\documentclass[11pt,twoside]{article}

\usepackage{asp2006}
\usepackage{epsf}
\usepackage{psfig}
\usepackage{lscape}
\usepackage{graphicx}

\newcommand\etal{{\it et al. }}
\newcommand\kms{km~s$^{-1}$~}
\newcommand\msun{$M_\odot$}
\newcommand\lsun{$L_\odot$}

\begin{document}
\title{The Void Phenomenon Revisited}   
\author{Riccardo Giovanelli}   
\affil{Cornell University, Dept. of Astronomy, Ithaca, NY 14853, USA}    

\begin{abstract} 
The Void Phenomenon consists in the apparent
discrepancy between the number of observed dwarf halos in cosmic voids
and that expected from CDM simulations. We approach the problem
considering the challenging prospects of detecting field dwarf systems 
with halo masses $\leq 10^9$ \msun, via their possible HI emission. 
A brief review of recent work is followed by preliminary results from 
the ALFALFA survey, which suggest the possibility, but not yet the proof, 
that such objects may have been already detected towards the outskirts
of the Local Group.
\end{abstract}


\section{The Void Problem: Where are the Dwarfs?}

More than 30 years ago it started to become apparent that bright galaxies 
are generally found in dense environments and that the vast majority of
cosmic space is devoid of them. The high density regions are highlighted
by clusters, arising in filamentary structures; regions between them that
are underdense in galaxies brighter than $\sim L^*$ by an order of magnitude 
or more fit spherical volumes of diameter often exceeding 10--20 Mpc, 
the ``voids''. This topology is now well
understood as a natural evolution of matter density fluctuations progressively 
amplified by gravitational instability. Current computer simulations readily
reproduce the main characteristics of the observed topology. Voids are not 
empty, either in simulations or in deep redshift surveys. Peebles
(2001, 2008) has however pointed out that a discrepancy appears to exist
between observations and simulations obtained within the standard Cold Dark Matter
(CDM) scenario: while a significant population of low mass halos is
predicted to exist by simulations, observations have failed to detect faint
galaxies in the expected abundance. An extension to voids of the 
morphology--density relation, seen to be relevant in higher density 
environments, does not seem to explain the discrepancy, as surveys of low
surface brightness (e.g. Thuan, Gott \& Schneider 1987) and blue compact dwarf 
galaxies (e.g. Salzer, Hanson \& Gavazzi 1990) replicate the topology of brighter
galaxies. Peebles has referred to this as the ``Void Phenomenon'', which appears
to be of analogous nature to the ``missing satellite'' problem pointed out by
Klypin \etal (1999) and Moore \etal (1999).

The organizers of this conference have asked me to revisit this issue, possibly
with the expectation that the partial results of the currently ongoing ALFALFA HI 
survey may be able to throw some new light, as that survey is highly sensitive to 
low mass, gas rich systems. In this report, an overview of key observations 
and high resolution simulations will be followed by results
which do not solve the problem --- if indeed a void problem still exists:
see the paper by Tinker in these proceedings and Tinker \& Conroy (2009) ---
but may provide interesting clues on how deep we need to go in order to detect
the baryonic counterparts of the halos that fill voids. Throughout this report 
I use $H_\circ=70$ \kms ~Mpc$^{-1}$.

\section{Simulations}

In its simulations, the {\it Mare Nostrum} collaboration has addressed with particular attention
the issue of dwarf galaxies in voids. In a pure DM simulation box of 115 Mpc size, 
Gottl\"ober \etal (2003) find that the 20 largest voids have radii larger than 25 Mpc, as
traced by $1.4\times 10^{12}$ \msun ~halos; when halos one order of magnitude smaller 
are used, the same voids are only 7\% smaller in linear size. The mean density within the volume of those 
voids is about 10\% of the cosmic density. Inside the voids, the same topology is found as
on larger scales in the Universe: empty regions, filaments and larger concentrations, but with masses
scaled down by orders of magnitude. If halo masses of $\sim 10^9$ \msun ~or smaller are used
to trace the topology, the voids are readily filled. Thus, the drop in the number density at the edge
of voids is quite steep for large halo masses and imperceptible for small ones. This translates in
a difference of an order of magnitude between the halo mass function of the general field and that
of large voids, at $10^{11}$ \msun; yet the two mass functions converge for $10^9$ \msun. Hence 
Peebles' query: where are the real world dwarfs? 

\begin{figure}[!th]
\hspace{1.0truein}
\includegraphics[height=3.in]{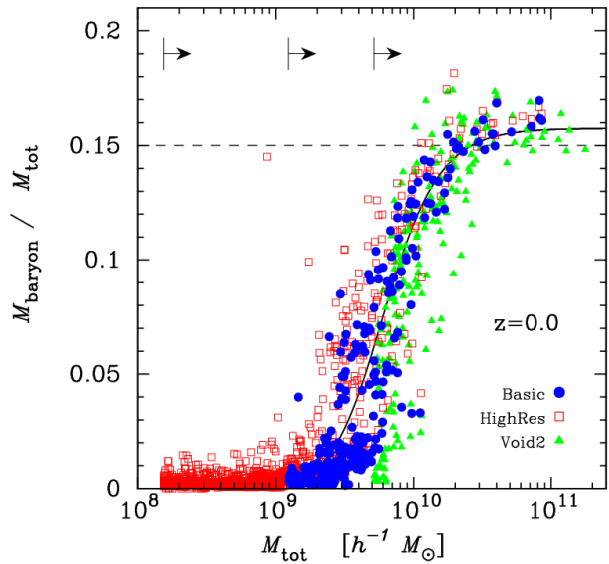}
\caption{Baryon fraction as a function of halo mass. Simulations by Hoeft 
\etal ~(2006) of different resolution: lowest are filled circles, highest 
are unfilled squares. The cosmic baryon fraction is 0.16. Included with
permission from M. Hoeft and the Royal Astronomical Society.} \label{hoeft.fig}
\end{figure}

Part of the answer is in Hoeft \etal (2006), who extended the earlier {\it Mare Nostrum} 
simulations with the inclusion 
of hydrodynamics and of a UV ionizing background. Their most important result is illustrated
in Figure \ref{hoeft.fig}, which shows the baryon fraction in halos of different masses, in 
various simulations of high resolution. While halos more massive than about $10^{10}$ \msun 
~are able to retain all their baryons, in those smaller than a few $10^9$ \msun ~most baryons
are heated by the intergalactic UV photons and lost by evaporation. While Hoeft \etal warn on
the possible inadequacies of their radiative transfer treatment, and while the relative absence 
of giant galaxies in voids attenuates the UV radiation flux, it is clear that the impact of the
latter and of other feedback processes on the ability of small halos to retain their baryons 
can be very important.
More recently, Ricotti (2009) has suggested that, as the Universe expands, variations in the 
Jeans mass in the IGM and increasing concentration of the halos can reactivate gas accretion
at late $z$, increase the minihalos' baryon fraction and stimulate late minibursts of star forming
activity.

A further source of uncertainty applies to the results of simulations, especially those
that adopt the so--called Halo Occupation Distribution (HOD) paradigm in populating DM 
halos with baryons and inferring simulated galaxy properties. According to Tinker \& Conroy (2009),
``the simple proposition within our implementation of the HOD is that galaxy properties are
determined solely by the mass of the halo in which the galaxy resides, independent of the
halo's larger scale environment". A number of authors have argued that an ``assembly bias"
exists, i.e. that the properties of DM halos and the galaxies embedded in them may depend
not only on their mass but also on their formation history, which is affected by the
environment (e.g. Gao, Springel \& White 2005; Wechsler \etal 2006), and that such
bias may be stronger in lower mass halos. This issue remains unclear; see 
Tinker's paper in these proceedings.
 
\section{Observations}

Adopting Mathis \& White (2001) conversion from halo mass to blue 
luminosity --- with mild extrapolation ---, galaxies inhabiting halos of mass $10^9$ 
\msun ~and lower would typically be fainter than
about -15. Hoeft \etal have however shown that the halo mass function cannot be 
translated into a luminosity function without full consideration of hydro, radiative
transfer and feedback effects: fainter magnitudes than -15 are obtained for void dwarfs
if the baryon fractions predicted by Hoeft \etal simulations
are converted via even generous baryon mass-to-light ratios. Such luminosities
are not effectively sampled through the nearest 100 Mpc by current redshift surveys.

In a detailed study, Hoyle \etal (2005) extracted two very rich samples of 
galaxies from the SDSS --- ``distant'' and ``nearby''; each sample was 
divided into two density regimes: ``void'' and ``wall'', via a nearest neighbor 
algorithm. Galaxies in the distant sample peak at $M_r\simeq -19.5$ and extend to
$M_r\simeq -17.5$; the nearby
sample peaks at $M_r\simeq -16.5$ and extends to $M_r\simeq -14.0$.
The luminosity functions of the void and wall populations 
differ in $\Phi^*$ by a factor of 7 and in $M*$ by 0.9 mag, yet they have similar
faint end slopes ($\alpha\simeq 1.19$). The galaxies in the void sample have halo
masses $>10^9$ \msun ~and reside near the
inner edges of voids. These results are in agreement with the expectations of 
simulations. 

Because field dwarf galaxies tend to be gas rich, with $M_{HI}/L$
typically increasing with decreasing luminosity $L$, HI
searches for gas rich dwarf halo dwellers have been carried out. Besides the
challenging demands of such experiments in terms of sensitivity and sky coverage,
a most important one is spectral resolution. A $10^9$ \msun ~halo translates into
a circular velocity of less than 20 \kms, thus spectral resolutions of 10 \kms or
better are needed in order to prevent signal dilution and allow reliable detection.
One of the most extensive HI searches, by Szomoru \etal (1996), covered $\sim 1$\%
of the Bootes void. Its spectral resolution, however, was $\sim 45$ \kms, allowing
detection of gas only in halos significantly more massive than $10^{10}$ \msun. This
result and those of other surveys have shown that such objects trace
voids in a similar manner as more massive ones, but little about the
extreme dwarf population.

\begin{figure}[]
\hspace{1.0truein}
\includegraphics[height=3.in]{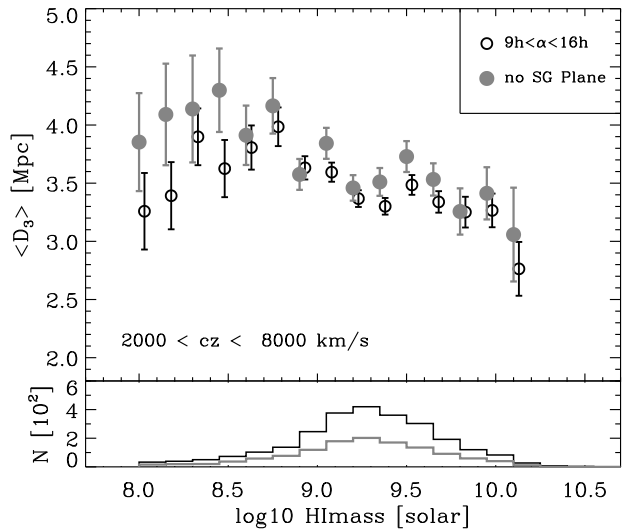}
\caption{Distance to the third nearest neighbor brighter than $M_r=-17.9$ (completeness limit
17.5 at cz=8000 \kms) vs. the HI mass in a preliminary subset
of the ALFALFA HI survey. This subset covers 900 contiguous square degrees (13\% of
the projected survey coverage) between $09^h$ and $16^h$, $4^\circ$
and $16^\circ$. Filled symbols refer to a subsample which excludes objects in the Supergalactic
plane (Saintonge \etal 2009).} \label{D3}
\end{figure}

The Arecibo Legacy ALFALFA extragalactic HI survey (Giovanelli \etal 2005), currently 
under way, aims to cover 7000 square degrees of sky, with a spectral resolution of 5.5 
\kms. As of Spring 2009, about 25\% of the survey data are fully processed. The survey
sensitivity allows detection of HI masses greater than $M_{HI}\geq 10^8$ \msun ~at 
distances $d>40$ Mpc, $2\times 10^7$ \msun ~at the distance of the Virgo Cluster (16.5 
Mpc) and $\sim 10^5$ \msun ~within the Local Group. Preliminary findings
of Amelie Saintonge \etal (2009), obtained from a 900 square degree contiguous region
fully sampled by ALFALFA, near the North Galactic Pole: $09^h<RA<16^h$, $4^\circ<Dec<16^\circ$,
are shown in Fig. \ref{D3}. The mean distance of HI detections to the third nearest
galaxy brighter than $M_r=-17.9$ is seen to change with $M_{HI}$: sources with $M_{HI}<10^{8.5}$
\msun ~inhabit environments less dense by about a factor of 3 than the mare massive
HI systems. This result confirms the report of Basilakos \etal (2007), based on much poorer
statistical grounds using data from the HIPASS survey. Completion of the ALFALFA survey
will permit extension of this result much further, both in distance and in mass regime,
than evidenced by Figure \ref{D3}.
In addition, over a solid angle of $\simeq 500$ square degrees containing part of the
volume of the giant void in the foreground of the Pisces--Perseus supercluster,
Saintonge \etal find that ALFALFA detects about a dozen HI sources with HI masses 
$>10^8$ \msun, 4 of which are optically brighter than -18. This is consistent with 
expectations from simulations. As argued in the next section, the sensitivity of ALFALFA
may not be sufficient to detect the possible cold baryonic counterpart of void dwarf 
halos with $<10^9$ \msun ~at distances of tens of Mpc, but it can allow testing the 
existence of such systems at smaller distances. 

\section{Are the Low Mass Halos Detectable?}

According to Fig. \ref{hoeft.fig}, a halo of $10^9$ \msun ~can retain a baryon mass of 
at most a few $10^7$ \msun. Wide field optical surveys such as 2MASS and SDSS
have been quite effective in discovering a significant number of  
dwarf satellites of the Milky Way. Their kinematics indicate total masses within 
a 300 pc radius on order of $2\times 10^7$ \msun, with $M/L>100$ and $10^3<L<10^5$ \lsun
~(Strigari \etal 2008). These faint, often tidally disrupted dwarf spheroidal systems are very 
nearby, largely within the virial radius of the MW itself and the total halo masses of 
their precursors are difficult to estimate. Their discovery is helping to alleviate 
the ``missing satellite'' problem. However, morphological segregation suggests that 
the typical field dwarf is more likely to be a gas rich, irregular system. This idea 
has been applied by Blitz \etal (1999) and Braun \& Burton (1999) to a category of
objects that has been known for decades: HI high velocity clouds (HVCs). They proposed 
that HVCs could be gas rich, optically impaired halos, spread over the Local Group. 
In that scenario, the so-called ``compact'' HVCs would be the relatively undisturbed
systems, while the very extended HVC complexes would be the tidally disturbed remnants
of ongoing infall to the Galactic disk. Catalogs of compact HVCs have been produced
(e.g. de Heij \etal 2002; Putman \etal 2002) and their global kinematics carefully
investigated. The typical dwarf halo associated with a compact HVC in this scenario
would have an observed HI mass of $10^7$ \msun, a linear size of 3-10 kpc and a total
mass in excess of $10^8$ \msun. The main objections to this idea arise from both
observations and theory. Observationally, systems with $M_{HI}\simeq 10^7$ \msun ~should 
have been detected in nearby groups of galaxies, other than the MW; they have not.
From the theory side, Sternberg \etal (2002) have shown (i) that most of the gas in a
minihalo should be found in a ionized phase, rather than in the form of HI, and (ii) that
the linear sizes of the compact HVCs, if located at typical LG distances ($\sim 1$ Mpc), 
are too large and would violate the halo mass-concentration relation. If that violation
were ignored, they also argue, the resulting halo models would require a baryon to DM 
mass fraction greater than the cosmic value of $\simeq 0.16$.

The models of Sternberg \etal (2002) provide useful templates for the thermal structure
minihalos would be expected to have. For example, a thermally stable dwarf system of 
halo mass $3\times 10^8$ \msun, embedded in a hot intergalactic medium of pressure 
$\simeq 10$ cm$^{-3}$ K, would have a gas mass of $1.8\times 10^7$ \msun, mostly warm 
($\simeq 10^4$ K) and ionized, envelopping a neutral core of $M_{HI}\simeq 3\times 10^5$
\msun, with a peak column density of $N_{HI}\simeq 4\times 10^{19}$ cm$^{-2}$ and mean
radius of the neutral gas $R_{HI}\simeq 0.7d$ kpc. The detection of such an object at
even modest extragalactic distance poses a stiff challenge, but it is possible with 
ALFALFA within distances of 2--3 Mpc. Objects with those characteristics were then 
searched for in the above--mentioned region near the NGP (Giovanelli \etal 2009). 
At high galactic latitude, intrusion of Galactic HI emission was minimized . About 
two dozen extremely compact sources were found, with $M_{HI}$ between $5\times 10^4 d^2$ 
and $10^6 d^2$ \msun, $R_{HI}$ between $0.4d$ and $2.8d$ kpc (several sources are
unresolved and smaller than 0.4 kpc), peak column densities
of a few $10^{19}$ cm$^{-2}$ and linewidths near 25 \kms, with the unknown 
distance $d$ in Mpc. A number of dwarf galaxies in the Local Group periphery, with
redshift independent estimated distances of 2.5 Mpc or less, are found in that 
region of the sky, including Sex B, GR8, KKH86 and DDO187. The velocity distribution 
of the newly discovered clouds matches well those of the nearby galaxies in the field.
Leo T, a dwarf at a distance of 0.42 Mpc, with $M_{HI}= 3\times 10^5$ \msun ~can
be said to be very gas-rich, as $M_{HI}/L_V\simeq 5$ (Ryan--Webber \etal 2008). 
With an HI radius $R_{HI}=0.3$ kpc and an optical counterpart that flies below the 
threshhold of optical surveys, the properties of Leo T are comparable with those
of the compact HI clouds discovered by ALFALFA. However, 
while those properties are consistent with the hypothesis that the new ALFALFA 
sources are the baryonic counterparts to a population of low
mass halos, it is not yet possible to exclude that they may be part of the wider
scenario of the (yet relatively poorly understood) perigalactic HVC phenomenon.
Completion of the ALFALFA survey will shed more light on this matter. The search
for analogs of these sources in nearby groups of galaxies and in
the Local (Tully) Void, well outlined by galaxies in the {\it Catalog of Nearby Galaxies}
of Karachentsev \etal (2004), is a near term, exciting prospect. 

This work was supported by the NSF grant AST-0607007.





\begin{thebibliography}{}
\bibitem[]{}
Basilakos, S. \etal 2007, \mnras ~378, 301

\bibitem[]{}
Blitz, L. \etal 1999, \apj ~514, 818

\bibitem[]{}
Braun, R. \& Burton, W.B. 1999, \aap ~341, 437

\bibitem[]{}
de Heij, V., Braun, R. \& Burton, W.B. 2002, \aap ~391, 159

\bibitem[]{}
Giovanelli, R. \etal 2005, \aj ~130, 2589

\bibitem[]{}
Giovanelli, R. \etal 2009, in preparation

\bibitem[]{}
Gottl\"ober, S. \etal 2003, \mnras ~344, 715 

\bibitem[]{}
Hoeft, M. \etal ~2006, \mnras ~371, 401

\bibitem[]{}
Hoyle \etal 2005, \apj ~620, 618

\bibitem[]{}
Karachentsev, I. \etal 2004, \aj ~127, 403

\bibitem[]{}
Klypin, A. \etal 1999, \apj ~522, 82

\bibitem[]{}
Mathis, H. \& White, S.D.M. 2002, \mnras ~337, 1193

\bibitem[]{}
Moore, B. \etal 1999, \apj ~524, L19

\bibitem[]{}
Peebles, P.J.E. 2001, \apj ~557, 495

\bibitem[]{}
Peebles, P.J.E. 2008, in {\it A Century of Cosmology}, p. 1035, 
ed. by Chincarini, G., Saracco, P. \& Bolzonella, M., Soc. Ital. 
di Fisica, Bologna

\bibitem[]{}
Putman, M.E. \etal 2002, \aj ~123, 873

\bibitem[]{}
Ryan--Weber, E.V. \etal 2008, \mnras ~384, 535

\bibitem[]{}
Saintonge, A. \etal 2009, in preparation

\bibitem[]{}
Sternberg, A, McKee, C.F. \& Wolfire, M.G. 2002, \apjs ~143, 419

\bibitem[]{}
Strigari \etal 2008, \nat ~454, 1096

\bibitem[]{}
Szomoru \etal 1996, \aj ~111, 2141

\bibitem[]{}
Tinker, J.L. \& Conroy, C. 2009, \apj ~691, 633

\end{thebibliography}
\end{document}